\documentclass[aps,twocolumn,nofootinbib,showpacs,tightenlines,showpacs,prd,amsmath,amssymb,nofootinbib,superscriptaddress]{revtex4}

\usepackage{latexsym}
\usepackage{graphicx}
\usepackage{subfigure}
\usepackage{hyperref}
\usepackage{cases}

\usepackage{color}

\newcommand{\beq}{\begin{equation}}
\newcommand{\eeq}{\end{equation}}
\newcommand{\beqa}{\begin{eqnarray}}
\newcommand{\eeqa}{\end{eqnarray}}

\def\nue{\mathrel{{\nu_e}}}
\def\numu{\mathrel{{\nu_\mu}}}
\def\nutau{\mathrel{{\nu_\tau}}}

\def \lta {\mathrel{\vcenter{\hbox{$<$}\nointerlineskip\hbox{$\sim$}}}}
\def \gta {\mathrel{\vcenter{\hbox{$>$}\nointerlineskip\hbox{$\sim$}}}}

\def\t13{\mathrel{{\theta_{13}}}}
\def\y12{\mathrel{{\tan^2 \theta_{12}}}}
\def\c2{\mathrel{{\chi^2 }}}

\newcommand{\n}{neutrino}
\newcommand{\ns}{neutrinos}

\newcommand{\oss}{oscillations}

\begin{document}

\title{Cosmic Strings as Emitters of Extremely High Energy Neutrinos}

\author{Cecilia Lunardini}
\email[]{Cecilia.Lunardini@asu.edu}
\affiliation{Physics Department, Arizona State University, Tempe, Arizona 85287, USA.}
\affiliation{RIKEN BNL Research Center, Brookhaven National Laboratory, Upton, New York 11973 USA.}
\author{Eray Sabancilar}
\email[]{Eray.Sabancilar@asu.edu}
\affiliation{Physics Department, Arizona State University, Tempe, Arizona 85287, USA.}

\begin{abstract}

\pacs{11.27.+d, 
98.70.Sa, 
      98.80.Cq 
}
We study massive particle radiation from cosmic string kinks, and its observability in extremely high energy neutrinos. In particular, we consider the emission of moduli --- weakly coupled scalar particles predicted in supersymmetric theories --- from the kinks of cosmic string loops. Since kinks move at the speed of light on strings, moduli are emitted  with large Lorentz factors, and eventually  decay into many pions and neutrinos via hadronic cascades.  The produced  neutrino flux has energy $E \gtrsim 10^{11}~\rm{GeV}$, and is affected by oscillations and absorption (resonant and non-resonant). It is observable at upcoming neutrino telescopes such as JEM-EUSO, and the radio telescopes LOFAR and SKA, for a range of values of the string tension, and of the mass and coupling constant of the moduli.
\end{abstract}

\maketitle


\section{Introduction}

Theories with spontaneous symmetry breaking usually have topologically non-trivial vacuum configurations. Depending on the topology of the vacuum after the symmetry breaking, stable relics called topological defects --- such as monopoles, strings or domain walls --- could be formed in the early universe \cite{Kibble:1976sj}. Strings can form if the vacuum manifold is not simply connected. 
Although monopoles and domain walls are generally problematic for cosmology, cosmic strings are compatible with the observed universe, provided that their tension is not too large (Sec.~\ref{general}; see, e.g., Refs.~\cite{VilenkinBook,Vachaspati:2006zz,Polchinski:2004ia,Copeland09,Copeland11} for reviews). 
 Cosmic strings are predicted in grand unified theories (GUTs) and superstring theory, and their existence can be revealed through their effects  on the cosmic microwave background (CMB), large scale structure and 21 cm line observations, and --- more directly --- by detecting their radiation, such as gravitational waves and cosmic rays. 

Since cosmic strings have GUT or superstring scale energy densities in their core, they can be significant sources of ultra high energy ($E \gtrsim 10^{11}$ GeV) cosmic rays \cite{Hill:1986mn,Bhattacharjee:1989vu,MacGibbon90,BOSV,Brandenberger:2009ia,Vachaspati10,BSV11}, either as isolated objects, or possibly in combination with
other topological defects, like in monopole-string bound states  \cite{Bhattacharjee:1994pk,Berezinsky:1997td,Berezinsky:1998ft,Berezinsky:1999az,BlancoPillado:2007zr}. Among the cosmic rays, neutrinos are especially interesting. Their weak coupling to matter makes them extremely penetrating, so they are the only form of radiation (together with gravitational waves) that can reach us from very early cosmological times, namely, all the way from redshift $z \sim 200$ (see Sec.~\ref{nuprop}). Moreover, in the spectral region of interest, $E \gta 10^{11}$ GeV,  the \n\ sky is very quiet, since this region  is   beyond the range of \ns\  from even the most extreme hadron accelerators (gamma ray bursts, supernova remnants, active galactic nuclei, etc.).  Therefore, even a low statistics \n\ signal beyond this energy would constitute a clean indication of a fundamentally different mechanism at play, such as a top-down scenario involving strings or other topological defects.  Experimentally, the technologies to detect ultra high energy \ns\ are mature: they look for radio or acoustic signals produced by the \ns\ as they propagate in air, water/ice, or rock. After the successful experiences of ANITA \cite{anita}, FORTE \cite{Lehtinen:2003xv}, RICE \cite{rice} and NuMoon \cite{numoon} --- the latter using radio waves from the lunar regolith via the so called Askaryan effect \cite{Askarian} --- a new generation of experiments is being planned, that can probe \ns\ from cosmic strings with unprecedented sensitivity. Of these, the space based fluorescent light telescope JEM-EUSO \cite{jem}, and radio telescopes LOFAR \cite{lofar} and SKA \cite{ska} seem especially promising. 

One of the distinguishing effects of cosmic strings as cosmic ray emitters is that they can produce bursts from localized features called cusps and kinks (Sec.~\ref{general}), where ultrarelativistic velocities are reached. The radiation from cusps and kinks is very efficient, whereas the emission from cusp/kink-free string segments is exponentially suppressed. This enhanced emission has been studied in connection with gravitational waves \cite{Garfinkle:1987yw,Garfinkle:1988yi,DV}, and electromagnetic radiation \cite{Vilenkin:1986zz,Garfinkle:1987yw,Garfinkle:1988yi,BlancoPillado:2000xy} like gamma ray bursts \cite{Paczynski,Berezinsky:2001cp,Cheng:2010ae} and radio transients \cite{Vachaspati:2008su,Cai:2011bi,Cai12}, as well as \n\ bursts  \cite{BOSV}.

Among the several scenarios considered, there are a few  that predict cosmic ray and \n\  fluxes at an observable level, e.g., Refs.~\cite{BOSV,Brandenberger:2009ia,Vachaspati10,BSV11}. One of these, Ref.~\cite{BSV11}, involves the decay of moduli --- massive scalar fields that arise in supersymmetric and superstring theories ---,  that can have various masses and couplings to matter. Moduli with coupling stronger than gravity are fairly natural \cite{Goldberger00,Conlon07,Frey06,Burgess10,Cicoli11} and relatively unconstrained due to their very short lifetimes \cite{Sabancilar10}, compared to gravitationally coupled ones \cite{Damour97,Peloso03,Babichev05}.  By decaying into hadrons, the moduli eventually generate a \n\ flux. In Ref.~\cite{BSV11} the emission of such moduli from string cusps, and the corresponding \n\ flux were discussed. 

In this paper, we elaborate on the theme of moduli-mediated \n\ production from strings, and  study modulus emission from kinks. We show that the emission from kinks is very efficient, and is the dominant energy loss mechanism for the cosmic string loops for a wide range of the parameters. We calculate the \n\ flux expected at Earth after a number of propagation effects, mainly absorption due to resonant ($Z^0$ resonance channel) and non resonant \n-\n\ scattering. We  find that the  flux might be observable at near future surveys, JEM-EUSO, LOFAR and SKA, depending on the parameters. 

The structure of the paper is as follows. After discussing some generalities on strings and kinks in Sec.~\ref{general}, the modulus emission from a cosmic string kink is calculated in Sec.~\ref{kinkrad}. In Sec.~\ref{particleprod}, we discuss the decay of moduli, the properties of the hadronic cascade initiated by their decay into gluons, and propagation of extremely high energy neutrinos in the universe. In Sec.~\ref{nuflux}, estimates are given for the kink event rate, the neutrino flux, and its detectability by the existing and future neutrino detectors. We also discuss the constraint from high energy gamma ray observations. Finally, in Sec.~\ref{disc}, we give our conclusions.


\section{Cosmic Strings}
\label{general}

Much of the phenomenology of a cosmic string depends on its {\it tension} (or mass per unit length), $\mu$. It  is often expressed in Planck units, as $G\mu$,  where $G$ is the Newton's constant.  Several cosmological and astrophysical observations place upper limits on $G\mu$;  we briefly review them here.

Since cosmic strings can create scalar, vector and tensor perturbations, they were initially considered as seeds for structure formation \cite{Zeldovich:1980gh,Vilenkin:1981iu,Silk:1984xk}. Hence, they contribute to the anisotropy of the cosmic microwave background (CMB) \cite{Kaiser:1984iv,Allen:1997ag,Albrecht:1997nt,Pogosian:2004ny,Wyman:2005tu,Fraisse:2007nu,Pogosian:2008am,Dvorkin:2011aj} and B-mode polarization of the CMB \cite{Seljak:1997ii,Bevis:2007qz,Pogosian:2007gi,Avgoustidis:2011ax}. However, the current measurements of the CMB anisotropies at small angular scales by WMAP \cite{Komatsu:2010fb} and SPT \cite{Keisler:2011aw}, reveal that the cosmic string contribution to the total power is less than $1.75\%$, which translates into a constraint on the string tension, $G\mu \lesssim 1.7\times 10^{-7}$ \cite{Dvorkin:2011aj}.

Although their contribution to the density perturbations is small, strings can still have effects on the early structure formation \cite{Shlaer:2012rj}, early reionization due to early structure formation \cite{Pogosian:2004mi,Olum:2006at}, formation of dark matter clumps \cite{Berezinsky:2011wf}, and might yield detectable signal in the 21 cm measurements \cite{Brandenberger:2010hn,Hernandez:2011ym,McDonough:2011er,Hernandez:2012qs,Pagano:2012cx}. Cosmic strings also produce gravitational waves \cite{Vachaspati:1984gt} in a wide range of frequencies, both as localized bursts and as stochastic background, which can be detected by LIGO, eLISA and pulsar timing array projects \cite{DV,pulsar,Olmez,Sanidas:2012ee,Dufaux12}. The most stringent bound comes from the pulsar timing measurements, which put an upper bound on the long wavelength stochastic gravitational wave background, $h^2\, \Omega_{\rm{GW}} \lesssim 5.6 \times 10^{-9}$ yielding the constraint $ G\mu \lesssim 4 \times 10^{-9}$ \cite{pulsar}. However, this upper bound is obtained by ignoring the kinetic energy of the cosmic string loops, and by assuming that cosmic strings only decay by emitting gravitational waves. Thus, the pulsar timing bound is expected to be somewhat relaxed by taking these effects into account. 

Cosmic string loops can emit moduli efficiently in the early universe when the length of the loop is of the order of the Compton wavelength of the emitted particle \cite{Damour97}. If moduli are gravitationally coupled to cosmic strings, very stringent cosmological constraints can be put on the string tension, $G\mu$, and the mass of the modulus, $m$ \cite{Damour97,Peloso03,Babichev05}. On the other hand, if their coupling is stronger than gravitational strength, modulus radiation becomes the dominant energy loss mechanism for the loops, and the lifetime of moduli becomes a lot shorter. These relax the cosmological constraints on moduli significantly \cite{Sabancilar10}. In this paper, we shall adopt the parameter space consistent with all the constraints mentioned above. 

Cosmic strings are born as smooth objects, but afterwards they undergo crossings and self crossings, which lead to truncations and successive reconnections. Every crossing produces a kink on the string after reconnection. The result of such processes are string loops with a few kinks \cite{Copi:2010jw}. Kinks are discontinuities in the vector tangent to the worldsheet characterizing the string motion, and gravitational and particle radiation is very efficient at kinks yielding waveforms with power law behavior in the momenta of the emitted particles \cite{Garfinkle:1987yw,DV}. There are also transient features on loops called cusps, where a part of the string doubles on itself, that reach the speed of light momentarily. Cusps also produce radiation in bursts, with waveforms that have a similar power law behavior. On the other hand, radiation from cosmic string loops with no cusps or kinks is exponentially suppressed, leaving the kink and cusp radiation as an interesting window on the observable effects from strings. In the next section, we shall study massive particle radiation from cosmic string kinks.


\section{Massive Particle Radiation from Kinks}
\label{kinkrad}

The free part of the action has Nambu-Goto term for a string of tension $\mu$, and the massive scalar field term for the modulus of mass $m$ 
\beq
S =  - \mu \int d^{2}\sigma \sqrt{-\gamma} - \int d^{4}x \sqrt{-g} \left( \frac{1}{2} |\partial_{\mu} \phi|^{2} + \frac{1}{2} m^{2} \phi^{2}\right ),
\eeq
where $g$ is the determinant of the spacetime metric $g_{\mu\nu}$ and $\gamma$ is the determinant of the induced metric on the worldsheet, $X^{\mu} (\sigma, \tau)$, given by $\gamma_{ab} = g_{\mu\nu} X_{,a}^{\mu}, X_{,b}^{\nu}$. The interaction Lagrangian for the modulus field and the string has the form \cite{BSV11}
\beq
\mathcal{L}_{int} = - \frac{\sqrt{4 \pi} \alpha \mu}{m_{p}} \phi \int d^{2}\sigma \sqrt{-\gamma},
\eeq
where $\alpha$ is the modulus coupling constant, $\mu$ is the string tension and $m_{p}$ is the Planck mass. Ignoring back reaction effects, the equation of motion for the worldsheet in the flat background $g_{\mu\nu} = \eta_{\mu\nu} = \rm{diag} (-1,1,1,1)$, and in the conformal gauge, where $\sigma^{0} = \tau$ and $\sigma^{1} = \sigma$, is 
\beq
\ddot{{\bf X}} - {\bf X}^{\prime \prime} = 0,
\eeq
with the gauge conditions $\dot{{\bf X}} \cdot {\bf X}^{\prime} = 0$ and $\dot{{\bf X}}^{2} + {\bf X}^{\prime 2} =1$. The general solution can be obtained in terms of the right and left moving waves as
\beq\label{X}
{\bf X}(\sigma, \tau) = \frac{1}{2} \left[ {\bf X}_{+} (\sigma_{+}) +  {\bf X}_{-} (\sigma_{-})\right], 
\eeq
where $\sigma _{\pm} \equiv \sigma \pm \tau$, and the gauge conditions are now given by 
\beq\label{gauge}
{\bf X}_{\pm}^{\prime 2} =1,
\eeq 
where the prime refers to the derivative with respect to the corresponding light cone coordinate $\sigma_{\pm}$.

The total power of particle radiation is
\beq
P = \sum_{n} P_{n},
\eeq
where the power spectrum $P_{n}$ can be calculated by using \cite{Damour97,WeinbergCosmoGrav}:
\beq\label{Pn}
\frac{dP_{n}}{d\Omega} = \frac{G \alpha^{2}}{ 2\pi} \omega_{n} k |T({\bf k}, \omega_{n})|^2,
\eeq
where $G = m_{p}^{-2}$ is the Newton's constant, $\alpha$ is the modulus coupling constant, $k$ is the momentum of the emitted particle, $\omega_{n} = 4\pi n/L = \sqrt{k^2 + m^{2}}$ is the energy, $L$ is the loop length, and $T({\bf k}, \omega_{n})$ is the Fourier transform of the trace of the energy momentum tensor of the cosmic string loop given by
\beqa\label{T}
T({\bf k}, \omega_{n}) &=& - \frac{4 \mu}{L} \int d^{4}x \nonumber \\
     &&
\times \int d\sigma d\tau \sqrt{-\gamma} \delta^{4} (x^{\mu} - X^{\mu} (\sigma, \tau)) e^{i k.x}.\,
\eeqa
Using Eq.~(\ref{X}) and the lightcone coordinates $\sigma_{\pm}$, Eq.~(\ref{T}) can be factorized as

\beq\label{T2}
T({\bf k}, \omega_{n}) = - \frac{\mu}{L} \int_{-L}^{L} d\sigma_{+} \int_{-L}^{L} d\sigma_{-} (1+ {\bf X_{+}^{\prime} \cdot X_{-}^{\prime}}) e^{\frac{i}{2}[\Phi_{+} -\Phi_{-}]},
\eeq
where $\Phi_{\pm} \equiv \omega_{n} \sigma_{\pm} \mp {\bf k\cdot X_{\pm}}$.

The integral in Eq.~(\ref{T2}) is exponentially suppressed\footnote{$L \leqslant 1/m$ condition is satisfied in the very early universe when the cosmic string loops can be quite small. Hence, moduli produced efficiently at earlier epochs can have cosmological effects, and subject to constraints \cite{Damour97,Peloso03,Babichev05,Sabancilar10}} for a smooth loop of cosmic string of length $L >> 1/m$ \cite{Damour97}. However, the phase become stationary if the string has cusps --- saddle points on the worldsheet where the derivative of the phase vanishes --- or kinks --- points where the vector tangent to the worldsheet has a discontinuity. The cusp case has been studied for massive particle emission in Ref.~\cite{BSV11}. At the cusp both $\Phi_{\pm}$ are saddle points, hence their derivatives with respect to the corresponding lightcone coordinates vanish. On the other hand, for the case of a kink, either $\Phi_{+}$ or $\Phi_{-}$ has a saddle point, and the other one has a discontinuity. In what follows we assume that $\Phi_{+}$ has a saddle point and $\Phi_{-}$ has a discontinuity. Then, assuming the kink is at $\sigma_{\pm} = 0$, worldsheet can be expanded about $\sigma_{\pm} = 0$ as follows:
\beq\label{X+}
{\bf X_{+}}(\sigma_{+}) \approx {\bf X}^{(0)}_{+} + {\bf X}^{(1)}_{+}\sigma_{+} + \frac{1}{2}{\bf X}^{(2)}_{+}\sigma_{+}^{2} + \frac{1}{6}{\bf X}^{(3)}_{+}\sigma_{+}^{3},
\eeq

\begin{numcases}
{{\bf X_{-}}(\sigma_{-}) \approx}
\sigma_{-} {\bf \hat{n}_{1}}  &  \qquad for  $\sigma_{-} < 0$\nonumber\\
 \sigma_{-} {\bf \hat{n}_{2}}  & \qquad for  $\sigma_{-} > 0$.\label{X-}
\end{numcases}

Using the gauge conditions (\ref{gauge}), one can show that ${\bf \hat{n}_{1}}$, ${\bf \hat{n}_{2}}$ and ${\bf X}^{(1)}_{+}$ are unit vectors, ${\bf X}^{(1)}_{+} \cdot {\bf X}^{(2)}_{+} = 0$ and ${\bf X}^{(1)}_{+} \cdot {\bf X}^{(3)}_{+} = - |{\bf X}^{(2)}_{+}|^{2}$. The curvature of the string can be approximated as $|{\bf X}^{(2)}_{+}| \sim 2\pi/L$ if the string is not too wiggly. Using the expansions (\ref{X+}) and (\ref{X-}), and the gauge conditions (\ref{gauge}), the phases $\Phi_{+}$ and $\Phi_{-}$ can be obtained as 
\beq\label{phi+}
\Phi_{+} \approx (\omega_{n} - k) \sigma_{+} + \frac{2 \pi^2}{3 L^{2}} k \sigma_{+}^{3},
\eeq
\begin{numcases}
{\Phi_{-} \approx}
\omega_{n} \sigma_{-} + k \sigma_{-} s_{1}  &  \qquad for  $\sigma_{-} < 0$\nonumber\\
\omega_{n} \sigma_{-} +   k  \sigma_{-} s_{2} & \qquad for  $\sigma_{-} > 0$,\label{phi-}
\end{numcases}
where $s_{1}$, $s_{2}$ are constants of order $1$, $|{\bf k}| \equiv k$ and we assumed that ${\bf k}~ //~ {\bf X}^{(1)}_{+}$. It can be shown that \cite{Vilenkin:1986zz,BlancoPillado:2000xy,DV} when moduli are emitted at a small angle rather than being parallel to the direction of ${\bf X_{+}}$ at the saddle point, the expansion (\ref{X+}) still applies provided that the angle satisfies 
\beq\label{theta}
\theta_{k} \lesssim (kL)^{-1/3},
\eeq
otherwise, leading to exponentially suppressed power. 

The term in the integrand of Eq.~(\ref{T2}) can be found as
\beq\label{integrand}
1+ {\bf X_{+}^{\prime} \cdot X_{-}^{\prime}} \approx c + \frac{c^{\prime}}{L} \sigma_{+} + O(\sigma_{+}^{2}),
\eeq
where $c$ and $c^{\prime}$ are constants of order $1$, which we will take as $1$ in what follows. Using Eq.~(\ref{integrand}), to the leading order we obtain
\beq
T({\bf k}, \omega_{n}) \approx - \frac{\mu}{L} I_{+} I_{-},
\eeq
where
\beq\label{Ipm}
I_{\pm} \equiv \int_{-L}^{L} d\sigma_{\pm} e^{\frac{i}{2}\Phi_{\pm}}.
\eeq
These integrals can be written explicitly by using Eqs.~(\ref{phi+}) and (\ref{phi-}) as follows
\beq
I_{+} = \int_{-\infty}^{\infty} d\sigma_{+} e^{\frac{i}{2}\left[(\omega_{n} - k) \sigma_{+} + \frac{2 \pi^2}{3 L^{2}} k \sigma_{+}^{3}\right]}~.
\eeq
After a change of variables, one obtains \cite{BSV11}
\beq
I_{+} = L\left(\frac{\omega_{n}}{k} -1\right)^{1/2} \int_{-\infty}^{\infty} dx e^{i\frac{3}{2} u\left(x + \frac{x^{3}}{3}\right)}, 
\eeq
where $u \equiv L k \left(\frac{\omega_{n}}{k} -1\right)^{3/2}$. The imaginary part of the integral vanishes, and the real part is given in terms of the modified Bessel function of order $1/3$
\beq
I_{+} = \frac{2}{\sqrt{3}} L\left(\frac{\omega_{n}}{k} -1\right)^{1/2} K_{1/3}(u).
\eeq
The function $K_{1/3}(u)$ exponentially dies out at large $u$, and it can be approximated as a power law in the limit $u<<1$ as $K_{1/3}(u) \approx u^{-1/3}$. This limit corresponds to $k>>m$, and in this regime, we can write $u \approx L m^{3}/16k^{2}$. Then, we obtain
\beq\label{I+}
I_{+} \sim L^{2/3} k^{-1/3},
\eeq 
where this formula is valid when $u \lesssim 1$, i.e., $k \gtrsim k_{c}$, where
\beq\label{kmin}
k_{c} \sim \frac{1}{4} m \sqrt{mL}.
\eeq
For smaller values of $k$, $I_{+}$ is exponentially suppressed, thus we are only interested in the above regime for practical purposes.

Using Eqs.~(\ref{phi-}) and (\ref{Ipm}), the integral $I_{-}$ can be similarly written as
\beq
I_{-} = \int_{-\infty}^{0} d\sigma_{-} e^{\frac{i}{2}[\omega_{n} \sigma_{-} + k \sigma_{-} s_{1}]} +  \int_{0}^{\infty} d\sigma_{-} e^{\frac{i}{2}[\omega_{n} \sigma_{-} + k \sigma_{-}s_{2}]} ~,
\eeq
which results in
\beq\label{I-}
I_{-} \sim \sqrt{\psi} k^{-1},
\eeq
where the sharpness of a kink is defined as 
\beq
\psi \equiv \frac{1}{2} (1- {\bf \hat{n}}_{1} \cdot {\bf \hat{n}}_{2}).
\eeq
Using Eqs.~(\ref{I+}) and (\ref{I-}), we find the power spectrum from Eq.~(\ref{Pn}) as
\beq\label{dpdk}
\frac{d^{2}P}{dk d\Omega} \sim \frac{\psi \alpha^{2} G \mu^{2}}{8\pi^2}  L^{1/3} k^{-2/3},~~~~ k \gtrsim \frac{1}{4} m \sqrt{mL}.
\eeq
Integrating over solid angle gives a factor 
\beq\label{Omegak}
\Omega_{k} \sim 2 \pi \theta_{k} \sim 2\pi (kL)^{-1/3},
\eeq
where $\theta_{k}$ given by Eq.~(\ref{theta}) is used. Then, the total power can be obtained as
\beq\label{Ptot}
P \sim \frac{\rm{ln}(\mu^{1/2}/m)^{3/2}}{4 \pi}  \psi \alpha^{2} G \mu^{2},
\eeq
where we used the cutoff for upper limit for the integral over momenta \cite{Vachaspati10,Olum99}
\beq\label{kmax}
k_{\rm{max}} \sim \mu^{3/4} L^{1/2},
\eeq
and the lower limit $k_{\rm{min}} \sim k_{c}$ from Eq.~(\ref{kmin}). Note that for typical values of the modulus mass $m$ and the string tension $\mu$, the logarithmic factor is about 20. Then, we can simply write the total power as
\beq\label{P}
P \sim \bar{\alpha}^{2} G \mu^{2},
\eeq
where we define $\bar{\alpha} \equiv \sqrt{\psi} \alpha$. Number of particles emitted from a kink with momenta $k$ in the interval $(k, k+dk)$ can be found from Eq.~(\ref{dpdk}) as
\beq\label{dNk}
dN(k) \sim \frac{dP(k) L}{k}  \sim \bar{\alpha}^2 G \mu^{2} L k^{-2} dk,~~~~ k \gtrsim \frac{1}{4} m \sqrt{mL}.
\eeq

In addition to moduli, cosmic string loops also produce gravitational radiation with the power \cite{VilenkinBook}
\beq
P_{g} \sim 50 G\mu^{2}.
\eeq
It is convenient to define the power as
\beq
P = \Gamma G\mu^{2},
\eeq
where
\begin{numcases}
{\Gamma \approx }
\bar{\alpha}^{2} & \qquad for $\bar{\alpha}^{2} \gtrsim 50$\nonumber \\
50 & \qquad for $\bar{\alpha}^{2} \lesssim 50$. \label{gamma}
\end{numcases}
The dominant energy loss mechanism for loops determines the lifetime of a loop as
\beq\label{tauL}
\tau_{L} \sim \frac{\mu L}{P} \sim \frac{L} {\Gamma G \mu}.
\eeq
Then, the minimum loop size that survives at cosmic time t is
\beq\label{Lmin}
L_{\rm{min}} \sim \Gamma G \mu t.
\eeq

In the next section, we shall discuss the decay of the moduli produced from cosmic string kinks into neutrinos via hadronic cascades, and the propagation of these neutrinos in the universe.


\section{Particle Decay and Propagation}
\label{particleprod}
For simplicity, throughout this paper we will assume a matter dominated flat universe model, which lets us carry out the calculations analytically. We assume cosmological constant $\Lambda = 0$, and the total density parameter $\Omega_{\rm{m}} + \Omega_{\rm{r}} = 1$ has matter and radiation components. We use the following values of the cosmological parameters: age of the universe $t_{0} = 4.4 \times 10^{17}~\rm{s}$, time of radiation-matter equality $t_{\rm{eq}} = 2.4 \times 10^{12}~\rm{s}$, $1+z_{\rm{eq}} = 3200$ \cite{Komatsu:2010fb}. The scale factor in the radiation and matter eras are respectively given by $a_{\rm{r}} \propto t^{1/2}$ and $a_{\rm{m}} \propto t^{2/3}$. Using $a/a_{0} = 1/(1+z)$, $a_{0} \equiv 0$, the cosmic time can be written in terms of redshift as $t = t_{0} (1+z_{\rm{eq}})^{1/2} (1+z)^{-2}$ in the radiation era, and $t = t_{0} (1+z)^{-3/2}$ in the matter era.


\subsection{Modulus decay}
\label{moddec}
The decay channel for moduli with the largest branching ratio is the decay into gauge bosons with the interaction of the form \cite{Conlon07}
\beq\label{bosonic}
\mathcal{L} \sim \frac{\alpha}{m_{p}} \phi F_{\mu\nu} F^{\mu\nu},
\eeq
for a modulus field $\phi$ and a gauge field of field strength $F_{\mu\nu}$. For the gauge bosons in the standard model, the modulus lifetime is estimated as
\beq
\tau \sim 8 \times 10^{-6} m_{4}^{-3} \alpha_{3}^{-2}~\rm{s}.
\eeq
Since most of the moduli are emitted from kinks with momenta $k\sim m \sqrt{mL}/4$ [because of the decreasing power law given by Eq.~(\ref{dNk})], their lifetime is boosted by a factor of $\gamma \sim \sqrt{mL}/4$. For the fiducial values of the parameters, the Lorentz factor of a modulus emitted at redshift $z$ and survive at present epoch is given by
\beq\label{gammaz}
\gamma(z) \sim \frac{\sqrt{m \Gamma G\mu t}}{4 (1+z)} \sim 6.5 \times 10^{13} \frac{\Gamma^{1/2} \mu_{-17}^{1/2} m_{4}^{1/2}} {(1+z)^{7/4}} \, , 
\eeq
where we have used the fact that loops of size $L_{\rm{min}}$ given by Eq.~(\ref{Lmin}) yield the dominant contribution to the observable events, the factor of $(1+z)$ in the denominator takes into account the redshifting of the energy of the moduli emitted at epoch $z$. Thus, the ratio of the lifetime of a modulus emitted at redshift $z \lesssim z_{\rm{eq}}$ and decaying at redshift $z_{\rm{d}}$ to the cosmic time at epoch $z_{\rm{d}}$ is
\beq
\frac{\tau(z, z_{\rm{d}})}{t(z_{\rm{d}})} \sim 10^{- 3} \frac{ \mu_{-17}^{1/2}}{\Gamma^{1/2} m_{4}^{5/2}} \frac{(1+z_{\rm{d}})^{5/2}}{(1+z)^{7/4}}\, .
\eeq
Note that $\Gamma \geqslant 50$ from Eq.~(\ref{gamma}) and $z_{\rm{d}} \leqslant z$. Hence, moduli will decay in the same epoch, $z_{\rm{d}} \lesssim z$, as they are produced. Therefore, we assume that all the moduli decay before they reach the Earth. 

The most efficient channel for neutrino production from modulus decays is the decay into gauge bosons. In particular, gluons decaying into hadrons produce neutrinos with the largest multiplicity \cite{BOSV,BSV11}. The interaction of a modulus with a gluon field is of the form (\ref{bosonic}), and the hadronic cascade from these gluons produces numerous pions of either sign, which eventually decay into neutrinos and antineutrinos. For both, we expect a flavor composition in the ratio $\numu : \nue : \nutau = 2 : 1 : 0$, from the pion decay chain. 

The number of neutrinos per unit energy can be found by using the Dokshitzer-Gribov-Lipatov-Altarelli-Parisi (DGLAP) method. Monte Carlo simulations for the hadronic decay of a very massive particle show a power law behavior in energy as $E^{-n}$ with $n=1.9$ \cite{DGLAP}. For simplicity, we approximate the index as $n \approx 2$. Then, the fragmentation function has the form \cite{BOSV,BSV11}
\beq\label{frag}
\frac{dN_{\nu}}{dE} \equiv \xi(E,z) \approx 0.05 \frac{k}{(1+z)E^{2}}\, ,
\eeq
where $k/(1+z)$ and $E$ are the modulus and neutrino energies at the present epoch respectively. Here $E_{\rm{min}} < E < E_{\rm{max}}$ \cite{BSV11}, where 
\beq\label{Emin}
E_{\rm{min}} \sim \epsilon\, \gamma(z) \sim 6.5 \times 10^{13} \frac{\Gamma^{1/2} \mu_{-17}^{1/2} m_{4}^{1/2} \epsilon_{\rm{GeV}}} {(1+z)^{7/4}}~\rm{GeV},
\eeq
and $E_{\rm{max}} \sim 0.1\, k$. We take $\epsilon_{\rm{GeV}} \equiv \epsilon/(1 \rm{GeV}) \sim 1$ \cite{BSV11}. Since the neutrino spectrum has the form $E^{-2}$, most of the neutrinos will have the energy $E \sim E_{\rm{min}}$. This introduces a lower bound on the redshift, below which no neutrinos are produced with a given energy $E \lesssim E_{\rm{min}}$. For our estimates, we are interested in energies $E\gtrsim 10^{11}$ GeV corresponding to the minimum redshift in the matter era
\beq\label{zmin}
z_{\rm{min}}(E) \sim 40\, \Gamma^{2/7}\, \mu_{-17}^{2/7}\, m_{4}^{2/7}\, E_{11}^{-4/7},
\eeq
where $E_{11} \equiv E/(10^{11}~\rm{GeV})$. Since the maximum redshift from which neutrinos can propagate to us is set by the neutrino horizon $z_{\nu} \sim 200$ (see Sec.~\ref{nuprop}), requiring $z_{\rm{min}}\lesssim z_{\nu}$, we have the constraint on the parameters
\beq\label{zminconst}
G\mu \lesssim 2.8\times 10^{-15}\, \Gamma^{-2}\, m_{4}^{-1}\, E_{11}^{2}.
\eeq


\subsection{Neutrino propagation}
\label{nuprop}

The \n\ flux at Earth is affected by a number of propagation effects: the redshift of energy, flavor oscillation, quantum decoherence and absorption.  The redshift of energy will be included as we carry out the flux calculation in the next sections;  the other effects, instead, warrant a separate discussion, which is the subject of this section. 

The \oss\ of very high energy \ns\ have been discussed in detail (see, e.g., \cite{Pakvasa:2007dc}).  Oscillations in vacuum are a good approximation, as the refraction potentials  due to the intergalactic gas and to the cosmological relic \n\ background  (which is assumed to be CP-symmetric here) are negligible \cite{Lunardini:2000swa}.
For the large propagation distances we consider, the flavor conversion probabilities are energy independent, as the energy-dependent oscillatory terms average out \cite{Eberle:2004ua}.  For our predicted initial flavor composition, $\numu : \nue : \nutau = 2 : 1 : 0$ (Sec.~\ref{moddec}), the effect of \oss\ is to equilibrate the flavors \cite{Pakvasa:2007dc}, therefore the composition at Earth should be $\numu : \nue : \nutau = 1 : 1 : 1$, for both \ns\ and antineutrinos. 

A neutrino oscillates as long as its wavepacket remains a coherent superposition of mass eigenstates. Depending on the size of the produced wavepacket, decoherence can occur as the \n\ propagates, due to the different propagation velocities of the mass states.  Dedicated analyses \cite{Eberle:2004ua,Farzan:2008eg} have shown that \ns\ of the energies of interest here remain coherent over cosmological distances, therefore we do not consider decoherence effects. 

Absorption effects  are largely dominated by scattering on the relic cosmological background \cite{Berezinsky92,Roulet:1992pz}, with negligible contribution from other background species.  In first approximation,  absorption can be modeled as a simple disappearance of the \n\ flux; secondary \ns\  generated by scattering are degraded in energy and therefore they are  negligible compared to primary flux.

The survival probability for the primary neutrinos, of observed energy $E$ and production redshift $z$, is defined as \cite{Berezinsky92,Weiler:1982qy,Roulet:1992pz,Barenboim:2004di}:
\beq
P(E,z) = e^{-\tau_{\nu}(E,z)}\ ,
\label{psurv}
\eeq
where the optical depth for the relic neutrino background is
\beq\label{optdepth}
\tau_{\nu} (E,z) = \int_{t(z)}^{t_{0}} dt^{\prime}\, \sigma_{\nu \nu}(E, \tilde{z})\, n_{\nu}(\tilde{z}),
\eeq
and
\beq
dt^{\prime} = - \frac{3}{2} \frac{d\tilde{z}}{(1+\tilde{z})^{5/2}},
\eeq
in the matter era. Here $n_{\nu}(z) =  56\,(1+z)^3~{\rm cm^{-3}}$ is the number density of relic neutrinos in each of the six species (\ns\ and antineutrinos of each flavor), and $\sigma_{\nu \nu}(E, z)$ is the neutrino-neutrino cross section, evaluated at the production energy $E^\prime = E(1+z)$, and summed over all the \n\ species in the background.
For the energies of interest here, and at the leading order, this cross section is the same for \ns\ and antineutrinos, and is practically flavor-independent: 
\begin{eqnarray}
\sigma_{\nu \nu } &= &\sigma_e \simeq \sigma_\mu \simeq \sigma_\tau    \\
\sigma_\alpha &=& \sum_{\beta = e,\mu,\tau} \left[ \sigma(\nu_\alpha + \nu_\beta \rightarrow any ) + \sigma(\nu_\alpha + \bar \nu_\beta \rightarrow any ) \right] \nonumber .  
\label{abslength}
\end{eqnarray}

In the limit of massless \ns, $m_{\nu} \approx 0$,  the Z$^0$-resonance effects can be ignored and the maximum cross section is attained at $E \gtrsim 10^{11}$ GeV \cite{BSV11}:
\beq\label{sigmamax}
\sigma_{\rm{max}} \approx  \frac{N}{\pi} G_{F}^{2} m_{\rm{W}}^{2},
\eeq
where $N \sim 10-15$, $G_{F} = 1.17\times 10^{-5}~\rm{GeV}^{2}$, and $m_{W} \simeq 80.39$ GeV. Using Eq.~(\ref{sigmamax}) in Eq.~(\ref{optdepth}), and requiring $\tau_{\nu} =1$ for absorption, the neutrino horizon --- the maximum redshift from which the neutrinos with energy $E$ can propagate to us --- is given by \cite{Berezinsky92,Roulet:1992pz}
\beq\label{znu}
z_{\nu} \sim 200,
\eeq
for energies $E \gtrsim 10^{11}$ GeV. In this regime, $P(E, z)$  can be approximated as a step function 
\beq\label{apxpsurv}
P(E, z) \approx 1- \Theta(z - z_{\nu}),
\eeq 
which becomes handy when estimating the neutrino flux analytically.

\begin{figure}
\includegraphics[width=0.48\textwidth]{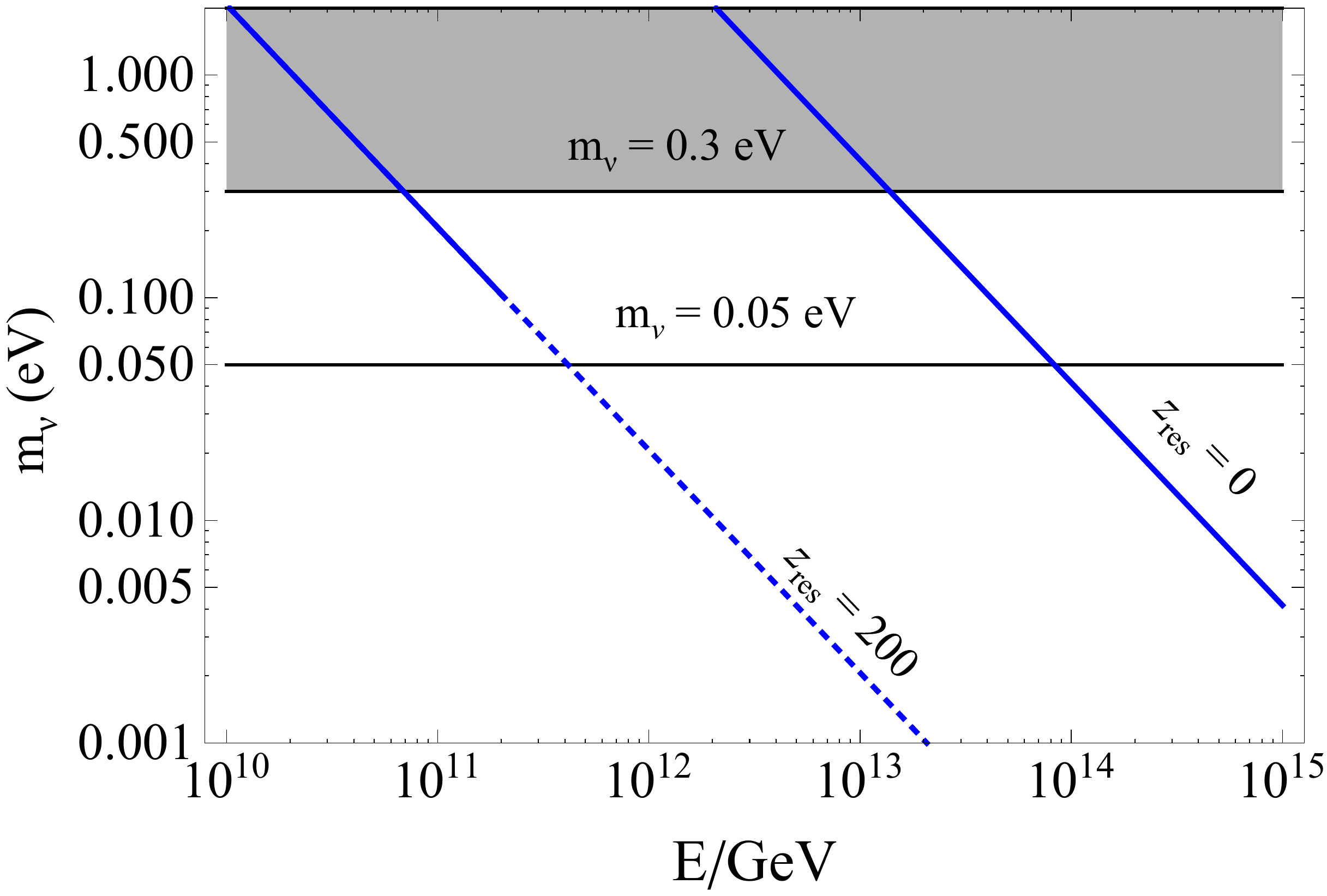}
\caption{ Graphical representation of the interval of observed energy, Eq.~(\ref{eabsint}), where suppression of the flux due to resonant $\nu-\bar\nu$ annihilation is expected (region between the diagonal lines).  The dashing indicates the region where Eq.~(\ref{eabsint}) is only indicative, due to thermal effects influencing the resonance.     The  horizontal shaded area refers to the interval of \n\ masses where the \n\ mass spectrum is strongly degenerate.  We also mark the value $m_\nu \simeq 0.05$ eV, which is the highest mass expected for non-degenerate (hierarchical) mass spectrum.   }
\label{zburstfig}
\end{figure}  

If the $Z^0 $ resonance is realized  in the annihilation channel $\nu_\alpha + \bar \nu_\alpha \rightarrow {\rm any}$, at a redshift $z_{\rm{res}}$ along the \n\ path, a pronounced dip in the \n\ spectrum is expected at the resonance energy due to the strong enhancement of the cross section \cite{Weiler:1982qy,Roulet:1992pz,Eberle:2004ua,Fargion:1997ft,Barenboim:2004di,D'Olivo:2005uh}: 
 \beq
 \sigma^{\rm{res}}_{\nu \nu} \propto \frac{ s}{  \left( s/M^2_Z -1 \right)^2 + \Gamma^2_Z/ M^2_Z}~,
 \label{zres}
 \eeq
 where $s\simeq 2 m_\nu E (1+z_{\rm res}) $ if the background \ns\ are not relativistic. The
   effect of the resonance is especially transparent in this case \cite{Weiler:1982qy,Roulet:1992pz}; we discuss it here in  its essentials.
  
Considering their  momentum, $p_\nu(z) \simeq 6.104 \times 10^{-4} (1+z) $ eV, the cosmological \ns\ are non-relativistic today for masses exceeding $\sim 10^{-3}$ eV, and throughout the interval of redshift of interest, $z \lta z_{\nu}$,  if $m_j \gg p_\nu(z_{\nu}) \simeq 0.1 $ eV.   From the data of oscillation experiments (see, e.g., \cite{Strumia:2006db}) it is known that, above this value, the \n\ mass spectrum becomes degenerate: $m_1 \simeq m_2 \simeq m_3 $.  Therefore, we can reason in terms of a single \n\ mass value, $m_\nu$, and take $m_\nu = 0.3$ eV as reference.  The degenerate case is optimal for the observability of the resonance effect, because the dip in the spectrum occurs at the same energy for all \ns\ and has a sharp shape.  Furthermore, it is located in the region of the spectrum, $\sim 10^{11}-10^{13}$ GeV,  where experiments have good sensitivity \cite{Eberle:2004ua} (see Fig.~\ref{zburstfig}). 

For a \n\ of energy $ E $ at Earth,  the  $Z^0$ resonance is realized at redshift $z_{\rm{res}} $ if 
\beq
E = \frac{M^2_Z}{2 m_\nu (1+z_{\rm{res}})}~ \simeq 1.4 \times 10^{13} {\rm GeV} \left( \frac{0.3 {\rm ~eV}}{m_\nu} \right) (1+z_{\rm{res}})^{-1}~,
\label{zres}
\eeq
with $M_Z \simeq 91.19$ GeV the mass of the $Z^0$ boson.  It follows that the flux of \ns\ of observed energy 
\beq
E =  \left( 6.9 \times 10^{10}  - 1.4 \times 10^{13}  \right) {\rm GeV} \left( \frac{0.3 {\rm ~eV}}{m_\nu} \right),
\label{eabsint}
\eeq
is affected by  the resonance between $z=z_{\nu}$ and the present epoch (see Fig.~\ref{zburstfig}), and therefore should be strongly suppressed compared to the flux at energies outside this interval, where the smaller, non-resonant, absorption cross section is at play. 

 Following the detailed discussion in Ref.~\cite{Barenboim:2004di}, we calculated $P(z,E)$ and used it to obtain the \n\ flux expected at Earth from all sources at all redshifts.  This flux is calculated by convolving the flux per unit of production redshift with the probability $P(z,E)$; it exhibits the characteristic suppression dip in the interval given in Eq.~(\ref{eabsint}), as expected (see Sec.~\ref{nufluxresults}). 

The absorption pattern is more complicated if the \n\ mass spectrum is not degenerate, i.e., $m_1 \lta  m_2 \ll m_3 \simeq 0.05$ eV (or, $m_3 \ll m_1 \lta  m_2  \simeq 0.05$ eV). For this configuration the probability $P(z,E)$ has three distinct dips of resonant suppressions at  separate resonance energies \cite{Eberle:2004ua,Barenboim:2004di}, corresponding to the three masses. These dips are broadened, in energy, by the integration over the production redshift, and, most importantly, by thermal effects, which are important for in this range of \n\ masses \cite{Eberle:2004ua,Barenboim:2004di,D'Olivo:2005uh}.   We postpone a discussion of these effects to a forthcoming publication \cite{usprep}.


\section{Neutrino Flux and Detection}
\label{nuflux}

As kinks move along a loop of cosmic string, they emit particles in a fan-like pattern, and scan a ribbon of solid angle $\Omega \sim 2\pi \theta_{k}$ [see Eq. (\ref{theta})]. Thus, one can analogously visualize the radiation from a kink as a source of light emitted from a lighthouse passing by. An observer who happens to be within the beam direction sees particles as a burst event provided that the flux is detectable. In this section, we make order of magnitude estimates for the event rate for bursts and neutrino flux, and compare it with the existing and future neutrino experiments.


\subsection{Loop distribution}
The distribution of cosmic string loops has been studied both analytically \cite{Rocha:2007ni,Polchinski07,Dubath08,Vanchurin11,Lorenz:2010sm} and in simulations \cite{Bennett90,Allen90,Hindmarsh97,Martins06,Ringeval07,Vanchurin06,Olum07,Shlaer10,Shlaer11}. Although there seems to be a consensus on the distribution of subhorizon size large loops, there is still not a good understanding of the small loop distribution. In what follows, we use the results from the latest simulation that has the largest dynamical range up to date for the evolution of the cosmic string network \cite{Shlaer11}, where it has been confirmed that the large loops form with size $\beta t$, where $\beta \sim 0.1$, and $t$ is the cosmic time at which the loop is chopped off the network of long cosmic strings. The density of long strings is $\rho \sim \zeta \mu/t^{2}$, with $\zeta \sim 16$. Using this framework, we can estimate the number density of loops of length $(L, L+dL)$ that are formed in the radiation era and still survive in the matter era as
\beq\label{loopdensity}
n(L, t) dL \sim p^{-1} \zeta (\beta t_{eq})^{1/2} t^{-2} L^{-5/2} dL,       
\eeq
where $\Gamma G \mu t \leqslant L \leqslant \beta t_{eq}$ and $p$ is the reconnection probability. There are also loops formed in the matter era, however, we have verified that their number density is negligible compared to the loops surviving from the radiation era given by Eq.~(\ref{loopdensity}). The dependence of loop density on reconnection probability has not been resolved yet, however, it is expected that the loop density increases for decreasing reconnection probability as discussed in Refs. \cite{DV,Sakellariadou05}. For ordinary cosmic strings, $p=1$, and it has been estimated as $10^{-3} \leqslant p \leqslant 1$ for cosmic F- and D-strings \cite{Jackson04}. Note that the most numerous loops have size of order $L_{\rm{min}} \sim \Gamma G\mu t $. As we shall see in Sec.~\ref{nuflux}, those will give the most dominant contribution to the observable effects, such as the diffuse neutrino flux. 

When a loop of size $\beta t_{eq}$ is formed, it will decay by the time [see Eq.~(\ref{tauL})]
\beq
t \sim \frac{\beta t_{eq}}{\Gamma G \mu}.
\eeq
This loop can survive until epoch $z < z_{\rm{eq}}$ provided that 
\beq
1+z \gtrsim 0.07\, \Gamma^{2/3} \mu_{-8}^{2/3}.
\eeq
where $\mu_{-8} \equiv G\mu/ 10^{-8}$, $\Gamma \gtrsim 50$ is given by Eq.~(\ref{gamma}), and we used $t = t_{0}\, (1+z)^{-3/2}$. Hence, we can conclude that even for the maximum $G\mu$ allowed by the current bounds, the loops can survive all the way to the very recent epochs from which we can get observable effects unless $\Gamma$ is too large.


\subsection{Burst rate}

The number of kink bursts per unit time can be estimated as \cite{BOSV,BSV11}
\beq\label{ddotN}
d\dot{N} = \frac{d\Omega}{4\pi} \frac{n(L,t) dL}{L (1+z)/2} dV(z),
\eeq
where $n(L,t) dL/(L/2)$ is the frequency of a kink event per physical volume per unit loop length, $L/2$ is the oscillation period of a loop of length $L$, $d\Omega/4\pi \sim \theta_{k}/2 $ is the probability that an observer lies within the solid angle of kink radiation, and $dV(z)$ is the physical volume in the interval $(z,z+dz)$ in the matter era, given by
\beq
dV(z) = 54 \pi t_{0}^{3} (1+z)^{-11/2} [(1+z)^{1/2} - 1]^{2} dz.
\eeq  
To find the total burst rate, we integrate Eq.~(\ref{ddotN}) over $L$ and $z$. Integral over $L$ is dominated by its lower limit $L_{\rm{min}}$ given by Eq.~(\ref{Lmin}), and the integral over redshift is dominated by its upper limit $z_{\nu} \sim 200$. Numerically, we obtain the total event rate as
\beq\label{Ndot}
\dot{N} \sim 1.6\times 10^{18} \, p^{-1}\, \Gamma^{-3}\, \mu_{-17}^{-3}\, m_{4}^{-1/2} ~\rm{yr^{-1}}.
\eeq
Remember that $\Gamma \gtrsim 50$ is given by Eq.~(\ref{gamma}). Since experiments run for a few years, the event rate should be at least $\sim 1$ per year to get observable events. Requiring $\dot{N}\gtrsim 1\, \rm{yr^{-1}}$ yields the constraint on the parameters
\beq\label{Ndotconst}
G \mu \lesssim 1.2 \times 10^{-11} p^{-1/3}\, \Gamma^{-1}\, m_{4}^{-1/6} \dot{N}_{\rm yr}^{-1/3}\, ,   
\eeq
where $\dot{N}_{\rm yr} \equiv \dot{N}/{\rm (1\, yr)}$.


\subsection{Neutrino flux}
\label{nufluxresults}

The diffuse neutrino flux is obtained using the flux  from a single kink on a loop, and summing over all the loops in a volume constrained by the neutrino horizon $z_{\nu} \sim 200$. It can be estimated by
\beq
J_{\nu}(E) = \int \frac{d\dot{N}}{\Omega_{k} r^{2}} \xi(E, z) dN(k) P(E, z),
\eeq
where 
\beq\label{rz}
r(z) = 3 t_{0} (1+z)^{-1/2} [(1+z)^{1/2} -1],
\eeq
is the physical distance to the source at redshift $z$ in the matter era, $d\dot{N}$ is the kink event rate defined by Eq.~(\ref{ddotN}) and $\Omega_{k}$ is the solid angle into which moduli are emitted given by Eq.~(\ref{Omegak}). Here $\xi(E,z)$ is the fragmentation function given by Eq.~(\ref{frag}), $dN(k)$ is the number of moduli emitted from a kink with momenta k in the interval $(k, k+dk)$ given by Eq.~(\ref{dNk}), and $P(E, z)$ is the survival probability of neutrinos defined in Eq.~(\ref{psurv}). Putting everything together, we obtain
\beqa\label{nufluxfull}
E^{2} J_{\nu}(E) \sim 0.05 p^{-1} \beta^{1/2} \zeta \left(\frac{t_{eq}}{t_{0}}\right)^{1/2} \bar{\alpha}^{2} (G\mu)^{2} m_{p}^{2} t_{0}^{-1/2} \nonumber \\
\times \int_{z_{\rm{min}}}^{\infty} dz\, \frac{P(E,z)}{(1+z)^{7/2}} \int_{L_{\rm{min}}} \frac{dL}{L^{5/2}} \int_{k_{\rm{min}}}^{k_{\rm{max}}} \frac{dk}{k}.\,\,
\eeqa

\begin{figure}
\includegraphics[width=0.48\textwidth]{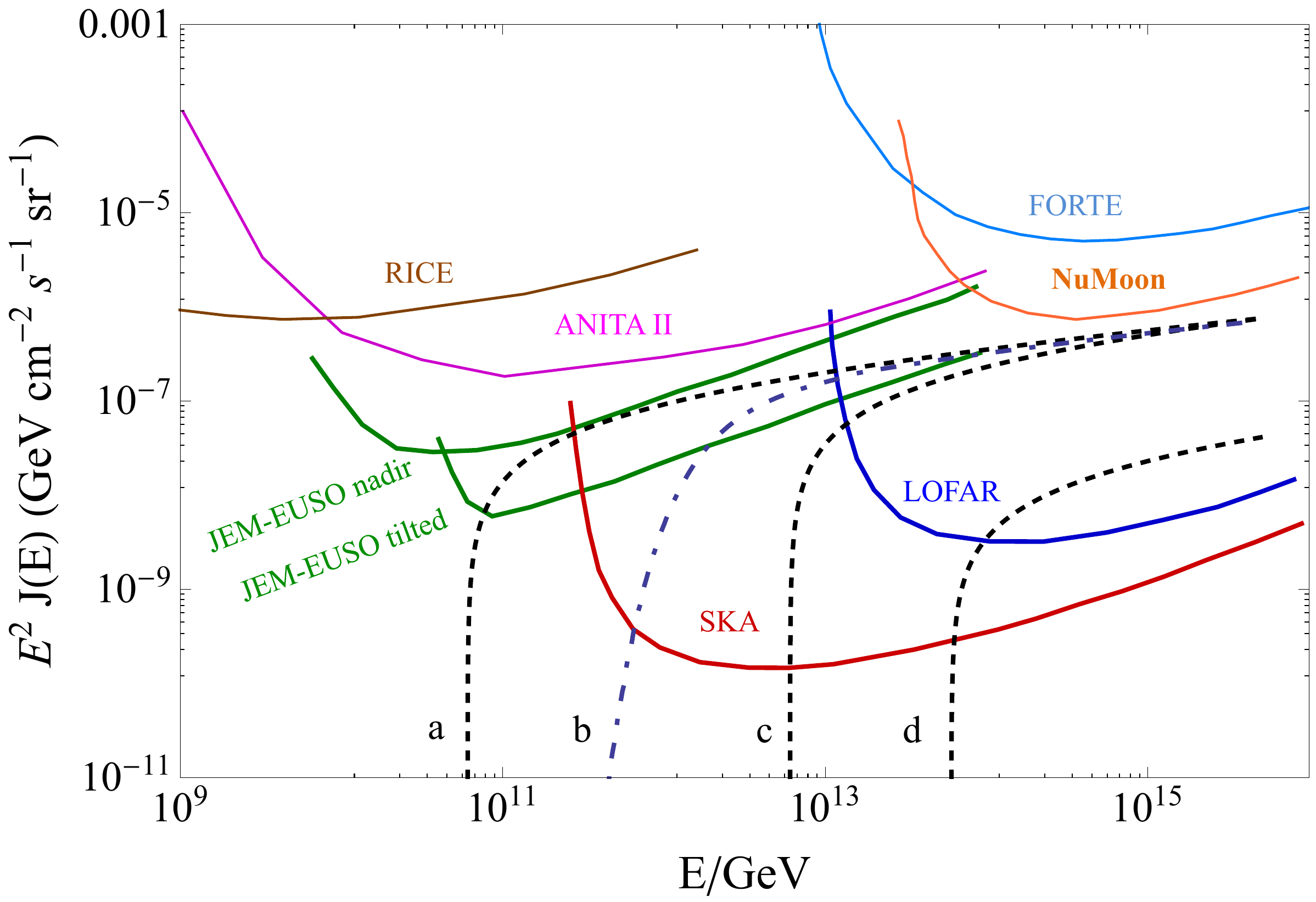}
\caption{Examples of \n\ flux from cosmic string kinks, via moduli decay.  We took $G\mu=10^{-17}$.  The dashed curves, (a),(c) and (d)  refer to $m_\nu = 0$ and $\bar \alpha = 1,10^2,10^3$ respectively.  The dot-dashed line, (b), is the same as (a) but for $m_\nu=0.3$ eV, with resonant absorption effects included. Cases (c) and (d) are not affected by the $Z^0$ resonance for this value of the \n\ mass.  The figure also shows existing limits from ANITA, FORTE, NuMoon and RICE \cite{anita,Lehtinen:2003xv,numoon,rice},  and expected sensitivities of the future detectors JEM-EUSO, LOFAR and SKA \cite{jem,lofar,ska}.  }
\label{detect}
\end{figure}

Note that the integral over $k$ gives a logarithmic factor $\rm{ln}(k_{\rm{max}}/k_{\rm{min}}) \sim \rm{ln}(\mu^{1/2}/m)^{3/2} \sim 20$ from Eqs.~(\ref{kmin}) and (\ref{kmax}), and the integral over $L$ is dominated by its lower bound $L_{\rm{min}}$ given by Eq.~(\ref{Lmin}). The integral over $z$ can be done numerically. However, it is useful to see the limiting case $m_{\nu} \approx 0$, where we can ignore the Z-resonance effects, and carry out the integral over redshift analytically. Using the approximate form of $P(E, z)$ given by Eq.~(\ref{apxpsurv}) in Eq.~(\ref{nufluxfull}), the neutrino flux can be calculated as
\beq
E^{2} J_{\nu}(E) \sim 2 \times 10^{-4}\, \frac{\mu_{-17}^{1/2} \bar{\alpha}^{2}}{p\, \Gamma^{3/2}}\,\left[z_{\rm{min}}^{-1/4} - z_{\nu}^{-1/4} \right] ~\rm{\frac{GeV}{cm^{2}\,s\,sr}}.
\eeq
Using $z_{\rm{min}}$ from Eq.~(\ref{zmin}), the predicted diffuse neutrino flux in the $m_{\nu} \approx 0$ limit is
\beq\label{flux}
E^{2} J_{\nu}(E) \sim 8 \times 10^{-5}\, \frac{\mu_{-17}^{3/7}\, \bar{\alpha}^{2}\, E_{11}^{1/7}}{p\, \Gamma^{11/7}}\, [1- (z_{\rm{min}}/z_{\nu})^{1/4}]~\rm{\frac{GeV}{cm^{2}\,s\,sr}}.
\eeq
Taking into account the neutrino mass in the survival probability $P(E,z)$, and fixing $m_{\nu} = 0.3$ eV and reconnection probability $p=1$, we evaluate Eq.~(\ref{nufluxfull}) numerically. In Fig.~\ref{detect}  we show the predicted flux for a few different values of the parameters $G \mu$ and $\bar{\alpha}$, together with the detectability limits of the current and future neutrino experiments.


\subsection{Diffuse gamma ray background constraint}

As moduli decay via hadronic cascades, the pions from this process also decay into photons and electrons. These high energy photons and electrons interact with the CMB photons and extra galactic background light, producing an electromagnetic cascade, whose energy density is constrained by the measurements of diffuse gamma ray background \cite{Berezinsky75}. The strongest upper bound on the cascade energy density comes from the highest energy end of the observed spectrum. The most recent data from Fermi-LAT observations reach $E \sim 100$ GeV \cite{fermilat}. The cascade photons with energy $E \gtrsim E_{\rm{abs}}$ will be strongly absorbed due to interaction with the CMB photos, where $E_{\rm{abs}}$ due to pair production can be estimated as
\beq
E_{\rm{abs}}(z) \sim \frac{m_{e}^{2}}{\epsilon_{\rm{CMB}} (1+z)} \sim 5.6 \times 10^{5} \frac{1}{1+z}~\rm{GeV},
\eeq
where $\epsilon_{\rm{CMB}} = 2.35 \times 10^{-4}$ eV and  $m_{e} = 0.511$ MeV. This implies that a cascade photon of energy above $E_{\rm{cas}} \sim 100\, (1+z)$ GeV, is efficiently absorbed at redshift 
\beq
1+z_{\rm{cas}} \sim \left(\frac{E_{\rm{abs}}(0)}{E_{\rm{cas}}(0)}\right)^{1/2} \sim 70.
\eeq

The electromagnetic energy density of the cascade from cosmic string kinks is \cite{BOSV,BSV11}
\beq
\omega_{\rm{cas}} = \frac{1}{2} f_{\pi} \int \frac{dt}{(1+z)^{4}} n(L,t)dL\, dP(k),
\eeq
where $f_{\pi}$ is the fraction of energy transferred to pions in the hadronic cascade initiated by a modulus decay, 1/2 is the fraction of energy transferred to electron-positrons and photons from the pion decays, $dP(k)$ is the power emitted from a kink, given by Eq.~(\ref{dpdk}). Integrating over $L$, $k$ and $z$ [similar to the diffuse flux in Eq.~(\ref{nufluxfull})], and integrating over $z$ up to $z_{\rm{cas}} \sim 70$, we have
\beq\label{cascade}
\omega_{\rm{cas}} \sim 1.2 \times 10^{-5}\,\frac{\mu_{-17}^{1/2} \bar{\alpha}^{2}}{p\, \Gamma^{3/2}}\,~\rm{\frac{eV}{cm^{3}}},
\eeq
where $\Gamma \gtrsim 50$ is given by Eq.~(\ref{gamma}). 

The maximum value of $\omega_{\rm{cas}}$ allowed by Fermi-LAT diffuse gamma ray data is $\omega_{\rm{cas}}^{\rm{max}} \sim 5.8 \times 10^{-7}\, \rm{eV/cm^{3}}$ \cite{Berezinsky11}. Therefore, $\omega_{\rm{cas}} \lesssim \omega_{\rm{cas}}^{\rm{max}}$ is satisfied for 
\beq\label{cascadebound}
G\mu \lesssim 3 \times 10^{-15} \bar{\alpha}^{-4} \left(\frac{\Gamma}{50}\right)^{3}.
\eeq
Note that $\omega_{\rm{cas}} \gtrsim \omega_{\rm{cas}}^{\rm{max}}$ is not strictly ruled out. This bound only constrains the observed highest energy diffuse gamma ray photons at $E_{\gamma} \sim 100$ GeV that are originated at very large redshifts $z \lesssim 70$. The constraints on the energy density of cascade photons produced at redshifts larger than $z_{\rm{cas}} \sim 70$ are much more weaker since they are more efficiently absorbed. Besides, the radiation from kinks is not homogenous, but confined to be in a narrow ribbon of width $2\pi \theta_{k} \ll 1$. Unless the cosmic magnetic fields are strong enough, the beamed electromagnetic radiation from cosmic string kinks might not diffuse efficiently, hence, the constraint might be relaxed significantly. Nevertheless, the examples given in Fig.~\ref{detect} respect the cascade upper bound given by Eq.~(\ref{cascadebound}).


\subsection{Neutrino bursts from individual kinks}

Before closing, we comment briefly on the possibility to identify the \n\ emission of individual kinks, i.e., bursts, rather than the diffuse flux.  The signature of a burst would be two or more time-coincident events  in  a detector\footnote{Due to the opening angle at emission, the coincident events would most likely appear in different regions of the detector field of view; this will allow to distinguish them from one another.}. Time coincidence at arrival is expected for \ns\ from a single burst, because  the emission occurs with a very short, practically vanishing, time scale of order \cite{Paczynski,Cai:2011bi}, 
\beq
\Delta \sim \frac{1}{k_{\min} \gamma^{2}} \sim 10^{-42}\, s,
\eeq 
for the fiducial values of the parameters, where $k_{\min}$ and $\gamma$ are respectively given by Eqs.~(\ref{kmin}) and (\ref{gammaz}). The time lag due to the spread in \n\ velocities is negligible for the  Lorentz factors of interest here, $\gamma \gtrsim 10^{11}$ [see Eq.~(\ref{Emin})]. 

The fluence of neutrinos with energy above $E$, from a kink on a cosmic string loop, can be estimated as \cite{BOSV,BSV11}
\beq
F_{\nu}(>E) = \int \frac{\xi(E, z)\, E\, dN(k) P(E, z)}{\Omega_{k} r^{2}}.
\eeq
Using Eqs.~(\ref{Omegak}), (\ref{dNk}), (\ref{frag}), (\ref{psurv}) and (\ref{rz}), and integrating over momenta yields
\beq
F_{\nu} (>E) \sim 4.3 \times 10^{-20}\, \frac{\bar{\alpha}^{2}\mu_{-16}^{15/4}\, \Gamma^{3/2} E_{11}^{-1}}{[\sqrt{1+z} -1]^{2} (1+z)^{9/4}}\, \rm{cm^{-2}},
\eeq 
where we take the loop length to be $L_{\min}$ given by Eq.~(\ref{Lmin}) since these loops are the most numerous, as was discussed in Sec.~\ref{nufluxresults}, hence it is more likely to get a burst from such loops. 

We can now estimate how many neutrinos might be detected at a detector of effective area
\beq
A_{\rm det} \sim  \sigma_{\nu N} \frac{M}{m_N} 
\eeq 
where the nucleon mass, $m_{N} \sim 1$ GeV,  $M$ is the target mass, and $\sigma_{\nu N} \sim 10^{-31}\, \rm{cm^{2}}$ is the \n-nucleon cross section. The reference cross section is from recent calculations at $E\sim 10^{12}$ GeV (see e.g., \cite{Goncalves:2010ay,Connolly:2011vc,CooperSarkar:2011pa}) and is a reasonable approximation for higher energies  as well, due to the slow rise of $\sigma_{\nu N}$ at these energies (less than $\propto E^{1/2}$). A typical value of the target mass for neutrino detection is $M \sim 10^{21}$ g, which applies to JEM-EUSO  in its nadir mode at energy $E \sim 10^{11}$ GeV \cite{jem}. 

We model a best case scenario by choosing the closest distance to the source, $z \sim z_{\min}$ (neutrinos can only come from $z_{\min}< z < z_{\nu}$), and the regime $\bar{\alpha}^{2} \gtrsim 50$ (where $\Gamma \sim \bar{\alpha}^{2}$), for which the  the number of emitted neutrinos is larger.

The number of events in a detector due to a burst can be estimated as:
\beqa\label{Nburst}
\mathcal{N}_{\nu} &\equiv& F_{\nu}(>E)\, A_{\det} \nonumber
\\
&\sim& 1.7 \, \bar{\alpha}_{3}^{3.14}\, \mu_{-17}^{2.82}\, E_{14}^{0.86} A_{16} ~, 
\eeqa
where $A_{16} \equiv A_{det}/(10^{16}\, \rm{cm^{2}})$. The fact that $\mathcal{N}_{\nu} \gtrsim 1$ means that, for our parameters of reference, the identification of a burst by time coincidence of multiple events is possible in principle, although in practice instrumental backgrounds might be an obstacle. 

Requiring $\mathcal{N}_{\nu} \gta 1$, implies a minimum value of $G\mu$: 
\beq
G\mu \gtrsim 7.8 \times 10^{-18} \bar{\alpha}_{3}^{-1.12} \, E_{14}^{-0.31} A_{16}^{-0.36}~.
\eeq
This has to be combined with the maximum value imposed by the condition that $z_{\min} \lesssim z_{\nu} \sim 200$ [see Eq.~(\ref{zminconst})] : 
$G \mu \lesssim 2.8 \times 10^{-15}\, \bar{\alpha}_{3}^{-4} m_{4}^{-1}\, E_{14}^{2}$.  The range of $G\mu$ is further restricted by imposing that burst detections are frequent enough, say one per year at least. Using  Eq.~(\ref{Ndotconst}), this gives $G \mu \lesssim 1.2 \times 10^{-17} \bar{\alpha}_{3}^{-2}\, m_{4}^{-1/6} \dot{N}_{\rm yr}^{-1/3}$. Therefore, the burst detectability is possible for a somewhat narrow range of $G \mu$.
 
Note also that a detector's capability to see bursts depends on its energy sensitivity: for most  of the parameter space, the \n\ emission is concentrated above the JEM-EUSO peak sensitivity, $E \sim 10^{11}$ GeV, therefore detection at JEM-EUSO might be hard. However, LOFAR and SKA are expected to surpass the JEM-EUSO sensitivity at higher energies (see Fig.~\ref{detect}), and therefore are more promising burst detectors.  The detection of a burst would be an important signature of cosmic string kinks or cusps (see Refs.~\cite{BOSV,BSV11} for bursts from cusps), complementary to a possible diffuse flux observation. It would also help breaking the degeneracy between the two parameters, $\bar{\alpha}$ and $G\mu$, since the detected number of neutrinos from a burst, (\ref{Nburst}), and the diffuse flux, (\ref{flux}), have different dependences on the parameters. Besides, even if only single neutrinos are detected, the rate of events, (\ref{Ndot}), can be used to help distinguish cosmic strings as the source, and break the degeneracy of the parameters.  


\section{Summary and discussion}
\label{disc}

Cosmic strings loops form as a result of reconnection of long strings, and self-intersection of large loops.  Kinks arise naturally as a result of these processes. We studied how kinks can radiate moduli, particles that arise in the supersymmetric models of particle physics, and that can have various masses and couplings to matter. The  decay of moduli into pions via hadronic cascades produces a flux of neutrinos, which can be observable depending on the parameters. 

Specifically, we considered the string tension $G \mu$, the modulus coupling constant $\alpha$ and mass $m$ as free parameters, and showed that neutrinos with energies $E \gtrsim 10^{11}$ GeV can be easily produced by cosmic string loops via this mechanism, with flux
\beq
E^{2} J_{\nu}(E) \sim 1.7 \times 10^{-7}\, \frac{\mu_{-17}^{3/7}\, \bar{\alpha}^{2}\, E_{11}^{1/7}}{p\, (\Gamma/50)^{11/7}}\,~\rm{\frac{GeV}{cm^{2}\,s\,sr}}.
\eeq
The hadronic cascade stops producing pions at the modulus rest frame energy of order $\epsilon \sim 1$ GeV. In the rest frame of the loop, this energy is boosted by the Lorentz factor $\gamma$, so that the
the minimum observed energy of the neutrinos is: 
\beq
E_{\rm{min}} (z) \sim 4.6 \times 10^{14} \frac{(\Gamma/50)^{1/2} \mu_{-17}^{1/2} m_{4}^{1/2} \epsilon_{\rm{GeV}}} {(1+z)^{7/4}}~\rm{GeV}. 
\eeq
The \n\ flux is shown in Fig.~\ref{detect} for representative sets of parameters; the termination of the flux at $E_{\min}$ appears clearly. The figure also gives the flux sensitivity of various experiments, showing that the predicted flux is within reach for the next generation neutrino detectors such as JEM-EUSO, LOFAR and SKA.

A distinctive feature of radiation from cosmic string kinks is that particles are emitted in a fan-like pattern, confined into a narrow ribbon, hence bursts from individual kinks can possibly be identified by timing and directional coincidence.  In Eq.~(\ref{Nburst}), we estimated the number of neutrinos emitted by a kink, and the corresponding number of events in a detector of a given effective area. We found that, for the fiducial values of the parameters used in our analysis, multiple neutrinos can be seen in the field of view of the detector. 

If ultra high energy \ns\ are observed at future experiments, what would be possible to learn? 
Top-down mechanisms would offer  natural explanations, and, among those, cosmic strings would be a favored candidate.  Even in the framework of cosmic strings, however, data analyses will necessarily be model-dependent, and various models would have to be considered.  Our scenario involving moduli is a possibility among many, and other intermediate states leading to \n\ production are possible, e.g., modulus emission from string cusps \cite{BSV11} and heavy scalar particle emission from cusps of superconducting strings \cite{BOSV}. Another possible generation mechanism of extremely high energy neutrinos could be the KK mode emission from cusps and kinks of cosmic F- and D-strings. The emission of KK modes of gravitons from cusps was studied in Refs.~\cite{DufauxKK1,DufauxKK2}, and various cosmological constraints have been put on the cosmic superstring tension. Depending on the parameters, observable neutrino fluxes might be produced by this mechanism as well.

A discrimination between different models will require the combination of complementary data, probably the detection of gravitational wave/electromagnetic counterparts of neutrino signals \cite{jointGWnu,jointGWem}.   The identification of point-like sources of extremely energetic \ns\ (bursts) would favor cosmic string kinks or cusps as sources, a hypothesis that would be substantiated further by the observation of accompanying gravitational wave and/or gamma ray bursts.  To distinguish between kinks and cusps could be possible since the event rate is larger for kinks for the given values of the parameters.  

In addition to a possible discovery of topological defects, detecting a flux of ultra high energy \ns\  might reveal new pieces of the still incomplete puzzle of \n\ physics.  Most interestingly, if the data show a $Z^0$ resonance dip, we might gather information on the \n\ mass and have another, perhaps more direct, evidence of the existence of the cosmological relic \ns.  The information on the \n\ mass might be especially important if at least one \n\ is light enough to evade a direct mass measurement in the laboratory. 

It is important to consider, however, that the extraction of any information from data would be complicated by many theoretical uncertainties. Let us comment on the uncertainties and  simplifying assumptions of our calculation.
First of all, we worked in a flat matter dominated universe, and ignored the recent accelerated expansion period of the universe, whose effect can be at most about a factor of a few in our final estimates. We also approximated the neutrino fragmentation function for the moduli decays as $dN/dE \propto E^{-n}$, and used $n=2$, whereas the numerical calculations yield $n \approx 1.9$ \cite{DGLAP}. In our estimates we take into account the reconnection probability $p$. For cosmic strings of superstring theory, namely, F- and D-strings \cite{Jackson04}, $p<<1$, whereas for ordinary field theory strings $p=1$. The flux, event rate and the chance of getting neutrino bursts is expected to be enhanced for cosmic superstrings with $p\lesssim 1$, compared to ordinary cosmic strings. We ignored the backreaction of modulus emission from kinks on the evolution of kinks. Since the total power from a kink is only logarithmically divergent [see Eq.~(\ref{Ptot})], the effect of radiation is expected to smooth out the sharpness of a kink slowly.   Finally, our treatment of the \n\ absorption due to resonant scattering on the \n\ backround is limited to relatively large masses, $m_{\nu} \gta 0.1$ eV, for which thermal effects on the background are negligible. The generalization to include these effects is forthcoming \cite{usprep}.


\acknowledgments
We would like to thank Veniamin Berezinsky, Jose Blanco-Pillado, Ken Olum, Ben Shlaer, Tanmay Vachaspati and Alex Vilenkin for useful discussions. This work was supported by the National Science Foundation grant No.~PHY-0854827 and by the Cosmology Initiative at Arizona State University.


\bibstyle{aps}

\end{document}